\documentclass[]{aa}

\usepackage{graphicx,amssymb,natbib}

\bibstyle{aa}
\bibpunct{(}{)}{;}{a}{}{,}

\begin{document}

\voffset=-1cm
\def\be{\begin{equation}}
\def\ee{\end{equation}}
\def\bea{\begin{eqnarray}}  
\def\eea{\end{eqnarray}}

\title{On non-linear hydrodynamic instability and enhanced transport in differentially rotating flows}

\author{ Denis  Richard \inst{1,2,3}}

\institute{ NASA Ames Research Center, MS 245-3, Moffett Field, California 94035
	\and  LUTH, Observatoire de Paris, F-92190 Meudon Cedex
	\and GIT-SPEC, Commissariat \`a l'Energie Atomique, l'Orme des merisiers, F-91190 Gif-sur-Yvette}

\offprints{D.Richard / \email{drichard@mail.arc.nasa.gov}}

\authorrunning{Richard}

\date{ Received ; Accepted June, 18 2003}

\abstract{ In this paper we argue that differential rotation can possibly sustain hydrodynamic turbulence in the absence of magnetic field. { We explain why
 the non-linearities of the hydrodynamic equations (i.e. turbulent diffusion) should not be neglected, either as a simplifying approximation or based on boundary counditions \citep{hbw}. The consequences of lifting this hypothesis are studied for the flow stability and the enhanced turbulent transport.
We develop a simple general model for the energetics of turbulent fluctuations in differentially rotating flows. By taking into account the non-linearities of the equations of motions, we give constraints on the mean flow properties for the possible development of shear instability}. The results from recent laboratory experiments on rotating flows show -- in agreement with the model -- that the pertinent parameter for stability appears to be the Rossby number $Ro$. The laboratory experiments seem to be compatible with  $Ro<-1/2$ and  $Ro > 1$ in the inviscid or high rotation rates limit. Our results, taken in the inviscid limit, are coherent with the classical linear stability analysis, in the sense that the critical perturbation equals zero on the marginal  linear stability curve.  We also propose a prescription for turbulent viscosity which generalize the $\beta$-prescription derived in \citet{RZ99}
}

\maketitle

\section{Introduction}

Differentially rotating flows are present in a wide variety of Astrophysical systems, including stellar interiors, accretion disks, or gaseous planets. Analytical studies teach us that differentially rotating flows are unstable, according to {\it linear stability theory}, whenever their angular momentum is decreasing outward \citep{Rayleigh}. These flows have been extensively studied but they do not exist in astrophysical context. Nevertheless, linear stability does {\it not} mean stability. This is because turbulence is genuinely a {\it non-linear} phenomenon. Even when the primary instability is linear in nature, the self-sustenance of fully-developed turbulence requires for the non-linearities to enter the game at least -- but not only -- for the saturation of amplitudes. In that sense, linear stability analysis is limited to predict, for a restricted class of flows, their instability and only the very early development of the bifurcated solution. It also means that linear analysis predicts instabilities but {\it does not} predict unconditional {stability}. Numerical models fail to maintain hydrodynamical turbulence in keplerian flows \citep{bhs}.  Based on past laboratory experiments from the 1930's \citep{Wendt,Taylor}, differentially rotating flows can become unstable for Reynolds numbers of order 10,000, a figure that to our knowledge has not been reached in published numerical simulations of keplerian flows.

\citet{RZ99} have shown, using laboratory experiments published in the 1930's, that differentially rotating flows exhibit finite-amplitude instabilities in the case where both angular velocity and angular momentum are increasing outward, a class of flows relevant for stellar interiors models. Recent laboratory studies \citep{Richard2002}, showed that the class of flows experiencing shear instabilities is much wider, as demonstrated by experimental results. We aim in this paper to find some hints about the physical mechanisms involved in these instabilities.


\section{Background and motivations}

 In this section we present the basic equations for the evolution of turbulent fluctuations. We discuss the relevant approximations that 
can be done and their implications regarding the properties of the turbulence and the mean flow.

   \subsection{Turbulent transport equations}
   
   We consider the equation of motion for a viscous incompressible flow,
   
\be \partial_{t} \vec{U} + (\vec{U} .  \vec{\nabla}) \vec{U} = - {1
\over \rho} \vec{\nabla} P + \nu \vec{\nabla}^{2} \vec{U}
\label{NS}
\ee
and we decompose the flow as its mean velocity and pressure fields and their fluctuations, namely
\be 
\vec{U}=\vec{u}+\vec{u^{'}} \hspace{1cm} P = p +p^{'}
\label{split}
\ee    
where $\vec{U}$ and $P$ are the total velocity and pressure fields, $\vec{u}$ and $p$ are the time averaged velocity and pressure and $\vec{u'}$ and $p^{'}$ are their time dependent fluctuations.  Multiplying the equations of motion by the fluctuation field and averaging over time, considering that
   
   \be
   \overline{\vec{u^{'}}}= \vec{0} \hspace{1cm} \overline{p^{'}} = 0,
\ee
(where $\overline{x}$ is the time average of $x$)
we find the evolution equations for the turbulent fluctuations kinetic energy (i.e. the diagonal components of the Reynolds stress tensor) reported in appendix \ref{ReynoldsStress}. Note that viscous diffusion has been neglected, but viscous dissipation has been kept. We consider circular motion  $\overline{u_{r}} = \overline{u_{z}} = 0$), with axial and azimuthal symmetries ($\partial_{z}= \partial_{\phi} = 0$) for the mean fields and define $\overline{u_{\phi} \over r } = \Omega $. The time averages of axial and azimuthal derivatives of fluctuating terms is neglected (i.e. net turbulent transport is only radial). 
\bea
 {1\over 2} \partial_{t} \overline{{u_{r}^{'}}^{2}} 
  =&& 2 \Omega \; \overline{u_{r}^{'}u_{\phi}^{'}} - \overline{{u_r^{'} \over r} \partial_r (r {u_r^{'}}^2)} + \overline{u_r^{'} {u_\phi^{'}}^2 \over r} \nonumber \\
&+& {1 \over \rho} \left( \overline{p^{'}\partial_r u_r^{'}} - \overline{\partial_r p^{'} u_r^{'}}\right) \nonumber \\
&-& \nu \left(  \overline{(\partial_r u^{'}_{r})^2} + {1 \over 2} \overline{(\partial_r u^{'}_{\phi})^2} + {1 \over 2} \overline{(\partial_r u^{'}_{z})^2 } \right),\label{ur}
\eea

\bea
{1\over 2} \partial_{t}\overline{{u_{\phi}^{'}}^{2}} =  &-& {\partial_{r} {\Omega r^{2}} \over r} \; {\overline{u_{r}^{'}u_{\phi}^{'}} } -\overline{u_\phi^{'}\partial_r ( u_r^{'} u_\phi^{'})} - \overline{{u_r^{'} u_\phi^{'}}^2  \over r}\nonumber \\
&-& \nu \left( {1 \over 2} \overline{(\partial_r u^{'}_{\phi})^2} + (\overline{{u^{'}_{r} / r}})^2  
+{1 \over 2} (\overline{{u^{'}_{\phi} / r}})^2 \right),\label{uphi}
\eea

\be
{1\over 2} \partial_{t}\overline{{u_{z}^{'}}^{2}} = - \overline{{u_z^{'}\over r} \partial_r (r  u_r^{'} u_z^{'})} - {\nu \over 2} \overline{(\partial_r u^{'}_{z})^2} .\label{uz}
\ee

\medskip

 This set of equations can be simplified by making the assumption that the turbulence is close to isotropy. This allows us to neglect the time averaged pressure fluctuations terms (see for example \citet{kato97}). We further assume that the characteristic spatial variation scale of the fluctuating components is smaller than the local radius (i.e terms of order $u^{'}_i u^{'}_j u^{'}_k / r$ can be neglected compared to the ones of order $\partial_r  u^{'}_i u^{'}_j u^{'}_k$). The above equations then become,

\bea
 {1\over 2} \partial_{t} \overline{{u_{r}^{'}}^{2}} 
  =&& 2 \Omega \; \overline{u_{r}^{'}u_{\phi}^{'}} - \overline{u_r^{'} \partial_r ( {u_r^{'}}^2)} \nonumber \\
  &-& \nu \left(  \overline{(\partial_r u^{'}_{r})^2} + {1 \over 2} \overline{(\partial_r u^{'}_{\phi})^2} + {1 \over 2} \overline{(\partial_r u^{'}_{z})^2 } \right)\label{ur2},
 \eea
 
 \be
{1\over 2} \partial_{t}\overline{{u_{\phi}^{'}}^{2}} =  - {\partial_{r} {\Omega r^{2}} \over r} \; {\overline{u_{r}^{'}u_{\phi}^{'}} } -\overline{u_\phi^{'} \partial_r ( u_r^{'} u_\phi^{'})} - {\nu \over 2} \overline{(\partial_r u^{'}_{\phi})^2} ,\label{uphi2}
\ee

\be
{1\over 2} \partial_{t}\overline{{u_{z}^{'}}^{2}} = -\overline{u_z^{'} \partial_r ( u_r^{'} u_z^{'})} - {\nu \over 2} \overline{(\partial_r u^{'}_{z})^2}.\label{uz2}
\ee

The total turbulent kinetic energy, defined as 
\be
k = {1\over 2}  \sum_{i=r,\phi,z} \overline{{u_{i}^{'}}^{2}},
\ee
then obeys the following equation
\be
\partial_t k = - \overline{u_{r}^{'}u_{\phi}^{'}} r \partial_r \Omega + \overline{u_i^{'} \partial_r u_r^{'} u_i^{'}} - \nu \overline{(\partial_i u_i^{'})^{2}} .\label{k}
\ee

 { The first term of the rhs in equations (\ref{ur2}) and (\ref{uphi2}) represents the production of turbulence. We will also refer to them as "coupling" terms, as they couple the turbulent fluctuations to the mean flow. The second term of the rhs in the same equations and the first one of equation (\ref{uz2}) describe the diffusion due to turbulence. The last term of equations (\ref{ur2})-(\ref{uz2}) is the viscous dissipation.}

\subsection{About linearized equations}

{ In this section we will be considering the effect of neglecting the non-linear terms (i.e. turbulent diffusion) in Equations (\ref{ur2})-(\ref{uz2}) and (\ref{k}).
There are several reasons why one would neglect the non-linearities of the hydrodynamic equations. The first one
is to simplify the system by neglecting terms that are difficult to treat in an analytical stability analysis. 
Other reasons might include the belief that the physical system is such that the non-linearities actually have a negligible effect
on its local or global evolution. Neglecting {\it a priori} such terms based on their local effect is rather difficult, considering
that little is know on their actual amplitude which does depend closely on the intrinsic nature of the turbulence. 
Another option is to consider spatially averaged equations and make some hypothesis on the boundary conditions of the systems. 
For accretion disks, it has been argue that the average over the whole flow or part of the flow of the turbulent diffusion terms eventually vanishes.
Making this hypothesis is equivalent to saying that there is no energy flux through the boundaries of the system. This is valid for an isolated system, but 
it seems unlikely for a disk, as it would not allow for energy transfer with its surroundings, in particular the central star. 
It does not seem realistic just by considering  the mass flux between the two objects. It is likely that such energy transfer occurs through the 
boundary layer between the star and the disk. Imposing that condition also formally implies that the turbulent stress tensor
vanishes on all boundaries, meaning that there is no turbulent viscosity at the edges of the disk. Which is more, the radial velocity also has to vanish
at the boundaries. We could also consider the properties of the boundaries beyond which the system can not be considered as a continuous medium, 
hence cease to be described accurately by the equations of hydrodynamics.  The description of the interactions between the fluid component 
of the disk and its outer non-continuous part is not a straightforward task.
Our last argument on this matter is to point that these turbulent diffusion terms are the ones that are modeled by the anomalous turbulent viscosity. 
Neglecting them is equivalent to making the assumption that there is no turbulent viscosity, no enhanced transport. It is then an inevitable result, 
that starting with such a truncated set of equations, one reach the conclusion that there is no such transport. 

Considering only the linear terms of the turbulent transport equations will give us insights --by definition-- only on the linear stability of the flow. 
The linearized set of equations is : }

\bea
 {1\over 2} \partial_{t} \overline{{u_{r}^{'}}^{2}} = && 2 \Omega \; \overline{u_{r}^{'}u_{\phi}^{'}} \nonumber\\
 &-& \nu \left(  \overline{(\partial_r u^{'}_{r})^2} + {1 \over 2} \overline{(\partial_r u^{'}_{\phi})^2} + {1 \over 2} \overline{(\partial_r u^{'}_{z})^2 } \right),\label{ur3}
\eea

\be
{1\over 2} \partial_{t}\overline{{u_{\phi}^{'}}^{2}} =  - {\partial_{r} {\Omega r^{2}} \over r} \; {\overline{u_{r}^{'}u_{\phi}^{'}} } - {\nu \over 2} \overline{(\partial_r u^{'}_{\phi})^2},\label{uphi3}
\ee
\be
{1\over 2} \partial_{t}\overline{{u_{z}^{'}}^{2}} = - {\nu \over 2} \overline{(\partial_r u^{'}_{z})^2},\label{uz3}
\ee
\be
\partial_t k = - \overline{u_{r}^{'}u_{\phi}^{'}} r \partial_r \Omega - \nu \overline{(\partial_i u_i^{'})^{2}}.\label{k3}
\ee

\medskip

{ From Eq. (\ref{k3}) we conclude that a stationary turbulent state ($\partial_t k = 0$) can exist only if the following condition --depending on the sign of the angular velocity gradient-- is satisfied}

\bea
\partial_{r} {\Omega } > 0 \;\; &\Rightarrow &\;\; {\overline{u_{r}^{'}u_{\phi}^{'}} } < 0 \; ,\nonumber \\
\partial_{r} {\Omega } < 0 \;\; &\Rightarrow &\;\; {\overline{u_{r}^{'}u_{\phi}^{'}} } > 0 \; .
\label{signeuruphi}
\eea

\medskip

For Rayleigh unstable flows (where $\partial_{r} {\Omega r^{2}} < 0$ and therefore $\partial \Omega <0$), eqs. (\ref{ur3}), (\ref{uphi3}) and (\ref{signeuruphi}) show that the linear coupling terms with the mean flow are always a source for turbulent fluctuations, a property that reflects the global linear instability of this class of flows. The non-linearities are needed there only for the saturation of amplitudes of the radial and azimuthal velocities along with the redistribution of energy towards axial motions of the new bifurcated flow. In the case of stable angular momentum stratification ($\partial_{r} {\Omega r^{2}} > 0$), the first order coupling terms have opposite signs in Eqs (\ref{ur2}) and (\ref{uphi2}). This means that one of them is an energy sink for one of the component of the velocity fluctuations \citep{bhs}.  It then looks impossible to allow for the growth or to maintain turbulence as long as the angular momentum is increasing outward. This actually reflects the linear stability of such flows.

\section{Considerations on the non-linear stability}

{ In the following sections,  will focus on the case of linearly stable flows (i.e. where $\partial_r \Omega r^2 >0$) and consider the energy equations including
the turbulent diffusion.}

\subsection{Inviscid flow}

{ We first consider the case where dissipation due to molecular viscosity can be neglected. We define $u$  the characteristic turbulent velocity (e.g. the root-mean-square velocity) and $\lambda$ the length scale  characteristic of the spatial variations of the velocity fluctuations, and we  pose}

\be
 {1\over 2} \partial_{t} \overline{{u_{r}^{'}}^{2}} 
  = 2 \Omega \; C^l_{r\phi} \; u^2 + C^{n}_r {u^3 \over \lambda} ,\label{ur4}
\ee
\be
{1\over 2} \partial_{t}\overline{{u_{\phi}^{'}}^{2}} =  - {\partial_{r} {\Omega r^{2}} \over r} \; C^l_{r\phi} \; u^2+ C^{n}_\phi {u^3 \over \lambda},\label{uphi4}
\ee
\be
{1\over 2} \partial_{t}\overline{{u_{z}^{'}}^{2}} = C^{n}_z {u^3 \over \lambda},\label{uz4}
\ee

\be
\partial_t k = -  r \partial_r \Omega \; C^l_{r\phi} u^2  + C^{n} {u^3 \over \lambda}.\label{k4}
\ee

\begin{table}
\caption{Correlation coefficients properties (Inviscid case)}
\begin{center}
\begin{tabular}{|c|c|c|c|c|c|}
\hline 
$\partial_r \Omega$ & $C_{r\phi}^l$ & $C_r^{n}$ & $C_\phi^{n}$ & $C_z^{n}$ & $C^{n}$\\
\hline$ >0 $ & $ <0 $ & $>0$ & $<0$ & $0$ & $<0$\\
\hline$ <0$ & $>0$ & $<0$ & $>0$ & $0$ & $<0$\\
\hline
\end{tabular}
\end{center}
\label{table1}
\end{table}

\medskip

{ We have introduced the correlation coefficients $C_r^{nl}$, $C_\phi^{n}$, $C_z^{n}$, $C_{r\phi}^l$ and $C^{n} = C_r^{n}+ C_\phi^{n} +C_z^{n}$. By imposing simple constraints, we can deduce the signs of these parameters, which depend on the sign of the angular velocity gradient. In the case of a negative gradient for example, we know from Eq. (\ref{k3}) that the correlation product $\overline{u_{r}^{'}u_{\phi}^{'}}$ has to be positive, which translates into $C^l_{r\phi}>0$. It also implies that a stationary state can be reach only if $C^{n}_r$ is positive and $C^{n}_\phi$ is negative. Note that  $C^{n}_z $  must be zero in that case because we neglected the viscous dissipation.} These properties for all introduced coefficients are summarized in Table (\ref{table1}).
From Eq. (\ref{k4}) we derive the amplitude of the fluctuations vorticity  :

\be
{u \over \lambda } = {C_{r\phi}^l \over C^{n}} r \partial_r \Omega \label{vort} \; .
\ee

This relation express that the vorticity extracted from the background flow by the fluctuations is proportional to the local shear. 

{ We have seen that there always exist a sink term for one of the components of the energy fluctuations. A necessary condition for the existence of self-sustained turbulence is then that the non-linear terms overcome this negative production term  (for the azimuthal component when $\partial_r \Omega <0$, for the radial one when $\partial_r \Omega <0$). Namely,  from Eqs (\ref{ur4}) and (\ref{uphi4})}

\bea
\partial_r \Omega > 0 \;\; : & & \;\;   C^{n}_r {u^3 \over \lambda} \geqslant  - 2 \Omega \; C^l_{r\phi} \; u^2 \;  \label{cond1a} , \\
\partial_r \Omega < 0 \;\; : & & \;\;   C^{n}_\phi {u^3 \over \lambda} \geqslant {\partial_{r} {\Omega r^{2}} \over r} \; C^l_{r\phi} \; u^2 \label{cond2a}\; ,
\eea

Which reduces to,

\bea
\partial_r \Omega > 0 \;\; : & & \;\;  {u \over \lambda} \geqslant \; - {C_{r\phi}^l \over C_r^{n}} \; 2 \Omega \;  \label{cond1} , \\
\partial_r \Omega < 0 \;\; : & & \;\;  {u \over \lambda} \geqslant \; {C_{r\phi}^l \over C_\phi^{n}}\; {\partial_r \Omega r^2 \over r}\label{cond2}\; ,
\eea

Finally, using Eq.(\ref{vort}), we obtain the following relations, 

\bea
\partial_r \Omega > 0 \;\; : & & \;\;  {C_{r\phi}^l \over C^{n}} r \partial_r \Omega   \geqslant \; - {C_{r\phi}^l \over C_r^{n}} \; 2 \Omega \;  \label{cond1c} , \\
\partial_r \Omega < 0 \;\; : & & \;\;  {C_{r\phi}^l \over C^{n}} r \partial_r \Omega   \geqslant \; {C_{r\phi}^l \over C_\phi^{n}}\; {\partial_r \Omega r^2 \over r}\label{cond2c}\; .
\eea

Introducing the dimensionless Rossby number, $Ro = r \partial_r \Omega / 2 \Omega$, and using the property $C^{n} = C_r^{n} + C_\phi^{n}$ we obtain the necessary conditions for instability :

\bea
\partial_r \Omega > 0 \;\; : & & \;\;  Ro  \geqslant  - {C^{n} \over C_r^{n}} = Ro_c \;\;\;\;\;\;\; ( Ro_c > 0 )\; , \\
\partial_r \Omega < 0 \;\; : & & \;\;  Ro  \leqslant    - {C^{n} \over C_r^{n}}= Ro_c \;\;\;\;\;\;\;  ( Ro_c < 0 )\;  \; .
\eea

{ Where we have introduced the parameter $Ro_c$. A similar result can be derived by simply posing $\partial_{t}<{u_{i}^{'}}^{2}> = 0$ and reducing equations (\ref{ur4}) and (\ref{uz4}), which leads to $Ro  =  - {C^{n} / C_r^{n}}$.}
The quantity $Ro_c$ can be seen as constant by assuming that the ratio $C_r^{n}/C^{n}$ is independent from the mean flow. This ratio quantify the redistribution of the energy extracted for the mean flow between the velocity components, by the non-linearities. Note that the coefficients $C_r^{n}$, $C_\phi^{n}$,$C_z^{n}$ and $C^{n}$ themself most likely depend on the mean flow (see \citet{speziale}). { Within this picture, $Ro_c$ is a constant critical stability parameter, and we expect the flow to become turbulent for either $Ro > Ro_c$ (when $\partial_r \Omega > 0$) or $Ro < Ro_c$ (when $\partial_r \Omega > 0$). }\citet{Townsend} already noted that the Rossby number should approach a constant value when a linearly stable rotating shear flow becomes turbulent, by arguing that "{\it it is conceivable and even likely that an asymptotic state can be reached with a constant ratio of transfer from the transverse motion to total energy transfer from the mean flow}".  { The two actors in the balance are the gradient of angular velocity, shearing the velocity fluctuations, creating small scale vorticity, feeding the turbulent cascade, and the gradient of angular momentum, stabilizing the flow by damping the fluctuations through a "spring" mechanism. The energy extracted from the background flow into small scale vorticity must reach a critical value in order to overcome the constraint introduced by the stable gradient of angular momentum. The Rossby number, as a measure of the ratio between the shear and the angular momemtum gradient, appears {\it a posteriori} as a natural control parameter of the flow stability.}

\subsection{Viscous flow}

\begin{figure*}
\centering
\includegraphics[width=15cm]{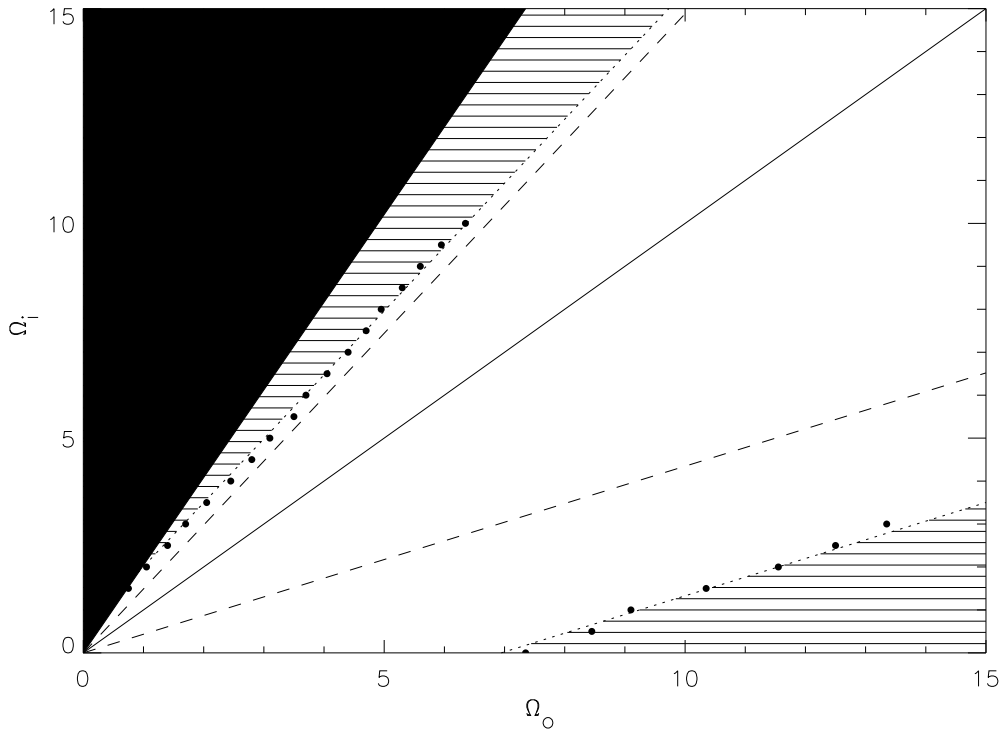}
\caption{\label{graph} Stability diagram of Couette-Taylor flow in the ($\Omega_o$,$\Omega_i$) plane where $\Omega_o$ and $\Omega_i$ are the angular velocities (in Hz) of the outer and inner cylinder respectively ;  solid line : solid body rotation ;  dotted lines : viscous stability criteria ; dashed lines : inviscid stability criteria (see text for detail). On the right (resp. left) of the solid body rotation curve, the angular velocity gradient is positive (resp. negative). The flows within the dark surface are Rayleigh unstable. The dots are the experimental instability boundaries within which the flows exhibit self-sustained turbulence ; The lined surfaces stand for the predicted non-linearly instability. Experimental results are from  \citet{Richard2002}}
\end{figure*}

\begin{table}
\caption{Correlation coefficients properties (Viscous case)}
\begin{center}
\begin{tabular}{|c|c|c|c|c|c|c|c|}
\hline 
$\partial_r \Omega$ & $C_{r\phi}^l$ & $C_r^{n}$ & $C_\phi^{n}$ & $C_z^{n}$ & $C^{n}$& $C^\nu_i$ & $C^\nu$\\
\hline$ >0 $ & $ <0 $ & $>0$ & $<0$ & $>0$ & $<0$&$>0$&$>0$\\
\hline$ <0$ & $>0$ & $<0$ & $>0$ & $>0$ & $<0$&$>0$&$>0$\\
\hline
\end{tabular}
\end{center}
\label{table2}
\end{table}

\medskip

Following the same path as in the inviscid case, we now add the viscous constraints to equations (\ref{ur4})-(\ref{k4}) :

\be
 {1\over 2} \partial_{t} \overline{{u_{r}^{'}}^{2}} = 2 \Omega \; C_{r\phi}^l \; u^2 + C_r^{n} {u^3 \over \lambda} - \nu C^{\nu}_{r} {u^2 \over \lambda^2} ,\label{ur5}
\ee
\be
{1\over 2} \partial_{t}\overline{{u_{\phi}^{'}}^{2}} =  - {\partial_{r} {\Omega r^{2}} \over r} \; C_{r\phi}^l \; u^2+ C_\phi^n {u^3 \over \lambda} - \nu C^{\nu}_{\phi} {u^2 \over \lambda^2},\label{uphi5}
\ee
\be
{1\over 2} \partial_{t}\overline{{u_{z}^{'}}^{2}} = C^{n}_z {u^3 \over \lambda} - \nu C^{\nu}_{z} {u^2 \over \lambda^2} ,\label{uz5}
\ee

\be
\partial_t k = -  r \partial_r \Omega \; C^l_{r\phi} u^2  + C^{n} {u^3 \over \lambda}.- \nu C^{\nu} {u^2 \over \lambda^2}\label{k5}
\ee

where $C^{\nu} =C^{\nu}_{r}+C^{\nu}_{\phi}+C^{\nu}_{z}$. The properties of the correlation factors are reported in Table (\ref{table2}). The extraction of vorticity from the mean flow now becomes :

\be
{u \over \lambda } = {C_{r\phi}^l \over C^{n}} r \partial_r \Omega   + {\nu \over \lambda^2} {C^\nu \over C^n} \label{vort2}  \; ,
\ee

and the relations (\ref{cond1}) and (\ref{cond2})  translate into

\bea
\partial_r \Omega > 0 \;\; : & & \;\;  {u \over \lambda} \geqslant \; - {C_{r\phi}^l \over C_r^n} \; 2 \Omega + {\nu C^{\nu}_{r} \over C_r^n \lambda^2}  \;  \label{cond1nu} , \\
\partial_r \Omega < 0 \;\; : & & \;\;  {u \over \lambda} \geqslant \; {C_{r\phi}^l \over C_\phi^n}\; {\partial_r \Omega r^2 \over r} + {\nu C^{\nu}_{\phi} \over C_\phi^n \lambda^2}\label{cond2nu}\; .
\eea

and after some rearrangement we finally obtain,

\bea
\partial_r \Omega > 0 \;\; : & & \;\;  Ro   \geqslant  Ro_c \left( 1-{1 / Re^{+}} \right)\; , \label{Ronu+}\\
\partial_r \Omega < 0 \;\; : & & \;\;  Ro   \leqslant   Ro_c \left( 1+{1 / Re^{-}} \right)\;  \; , \label{Ronu-}
\eea
where we have introduced the following quantities,

\be
Re^+ = { 2 \Omega  \lambda^2 \over \nu } (C_r^\nu /C_{r \phi}^l -{C^\nu C_r^n / C^n C_{r \phi}^l})^{-1} < 0 ,
\ee
  
\be
Re^- = { 2\Omega \lambda^2 \over \nu } ( C_\phi^\nu / C_{r \phi}^l-{C^\nu  C_\phi^n / C^n C_{r \phi}^l})^{-1} > 0 .
\ee
  
The correction due to viscous dissipation lies in the $1 / Re^{+}$ and $1 / Re^{-}$ terms. The necessary condition for instability reduces to the results found for the inviscid case, if  $\nu$ is set to zero. For high rotation rates, the critical values of the Rossby number will eventually tend to the inviscid values.

{ Considering all of the above, we expect that there exists a critical Rossby number for each class of flows -- namely flows with $\partial_r \Omega > 0$ and flows with $\partial_r \Omega < 0$.
This result can be verified by using experimental results on differentially rotating flows. This can be done by examining the value of the Rossby number at the onset of shear turbulence observed in laboratory Couette-Taylor experiment. The Couette-Taylor experiment consist of two coaxial cylinder between which the fluid is sheared, resulting in a differentially rotating azimuthal flow. Experimental results on stability \citep{Richard2002} can actually be fitted with the values  $Ro_c=1$ for $\partial_r \Omega > 0$, $Ro_c = -1/2$, for $\partial_r \Omega > 0$, $Re^+=-\Omega / 3$ and $Re^-=2 \Omega$. This is consistent with the picture of a constant critical Rossby number with a viscous correction vanishing for high rotation rates or low viscosity.}

\section{Turbulent transport and viscosity}

Radial turbulent transport of angular momentum is quantified by the second-order correlation product $\overline{u_{r}^{'}u_{\phi}^{'}}$. The time evolution for this
quantity after some rearrangement and spatial averaging over the azimuthal direction $\phi$, and the introduction of correlation coefficients, is given by  :

\be
\partial_t \overline{u_{r}^{'}u_{\phi}^{'}} 
=- C_{r\phi}^l u^2 r \partial_r \Omega + C_{r \phi}^n {u^{3} \over \lambda} - \nu C_{r \phi}^\nu {u^2 \over \lambda^2} + \Pi_{r\phi},
\label{uruphi}
\ee
where $\Pi_{ij}$ is the pressure-strain correlation tensor. Following \citet{kato97}, in the case of isotropic turbulence, and with their notations, we have 

\be
\Pi_{r\phi} = \frac{3}{2} C_2 {u^{'}}^2 r \partial_r \Omega .
\ee

We can derive a formal expression for the turbulent diffusion of momentum, using the classical definition ,

\be
\nu_t = {\overline{u_{r}^{'}u_{\phi}^{'}} \over r \partial_r \Omega} = {C_{r\phi}^l u^2 \over r \partial_r \Omega},
\ee

For steady turbulence,  in the limit where molecular viscosity in negligible compared to turbulent transport, from equation (\ref{uruphi}), 

\be
{\nu_t \over \nu} =  C_{r\phi}^l \left({{C_{r\phi}^l - (3/2) C_2 } \over {C_{r\phi}^{n}}}\right)^2  \left({\lambda \over r}\right)^2  { r^3 \partial_r \Omega\over \nu}
\label{nutf}.
\ee

 \citet{RZ99}  derived  an expression for the turbulent viscosity from \citet{Taylor} and \citet{Wendt} experimental data,

\be
{\nu_t \over \nu} = \beta \; { \big| r^3 \partial_r \Omega \big|\over \nu} = \beta Re^*
\ee 
where $\beta$ was found to be independent from the geometry and the background flow (see also \citet{Longaretti} and \citet{Duschl} for other arguments in favor of this prescription). Identifying this expression with Eq. (\ref{nutf}), one finds,

\be
\beta =\Big| C_{r\phi}^l \left({{C_{r\phi}^l - (3/2) C_2 } \over {C_{r\phi}^{n}}}\right)^2  \left({\lambda \over r}\right)^2\Big| .
\ee

 \citet{RZ99}  also identified the $Re^*$ parameter (named "gradient Reynolds number") as the control parameter for stability. This derivation  was acheived by identifying the values of the Reynolds number for which the measured torque from laboratory experiments starts to differ from its theoretical value for the laminar Couette flow. For small gap Couette-Taylor apparatus, the classical Reynolds number defined as $Re = \Delta \Omega \Delta R R / \nu$ (where $\Delta R$ is the gap between the cylinders, $\Delta \Omega$ their differential angular velocity, and $R$ the mean radius) was identified as the relevant control parameter. For gaps bigger than $\Delta R / R \gtrsim 1/20$ the critical Reynolds number behave like $(\Delta R / R )^2$, which means that $Re^*$ takes a constant value. To be more proper, this is a measure of the efficiency of the turbulent transport. The bifurcation identified from the torque data is then not directly linked to the stability of the flow, but is the threshold for which turbulent transport becomes appreciably more efficient than molecular viscosity. It should be read as the following condition 

\be
{\nu_t \over \nu} >> 1  \Rightarrow  Re^* >> {1 \over \beta}
\ee 

This is consistent with the numerical values from \citet{RZ99} ($Re^*_c = 1.5 \cdot 10^{-5}$ and $\beta^{-1} = 1.5 \cdot 10^{-6}$)

\section{Geometrical constraints}


\subsection{General case}

In the case of a bounded flow, keeping up the assumption of isotropic turbulence, we have to assume that $\lambda$ will be of order of a macro-scale $d$. The stability criteria for a viscous flow is modified as $\lambda$ is constrained to smaller values than for a free shear flow (therefore being closer to the inviscid values $Ro^+$ and $Ro^-$, according to Eqs (\ref{Ronu+}) and (\ref{Ronu-}).) while for the inviscid case it remains unchanged. The radial transport will also be modified, as we have, according to eq. (\ref{nutf})

\be
{\nu_t \over \nu} \propto  d^2  { | r {\partial_r \Omega}|\over \nu}
\label{nutf2}.
\ee

\subsection{Narrow gap Couette-Taylor experiment}

This expression can be approximated in the case of narrow-gap Couette-Taylor experiment by taking for $d$ the distance between the two cylinders $\Delta R$.

\be
{\nu_t \over \nu} \propto  {\Delta \Omega \cdot \Delta R \cdot R \over \nu} = Re
\label{nutf3},
\ee

The bifurcation threshold identified on torque measurement aforementioned will then scale as the classical Reynolds number, in agreement with the results from \citet{RZ99}.

\section{Discussion}

We have shown that a simple model of turbulent fluctuations energetics can explain laboratory experiment results on differentially rotating flow.
The equations of motion show that shear instabilities can develop in rotating flows, as long as the non-linearities are taken
into account. Making reasonable assumptions about the necessary properties of the velocity correlation products, we derived formal expressions
for the control parameters for stability. Comparison with laboratory results confirms that the relevant parameter is the Rossby number. In the case of a viscous flow, the critical Rossby numbers exhibit a $\Omega^{-1}$ correction, therefore converging toward the same constant
critical Rossby numbers as in the inviscid case, for large rotation rates. The turbulent viscosity inferred is consistent with the prescription previously proposed by \citet{RZ99}. The critical amplitude of perturbations are also consistent with linear stability theory results in the inviscid limit. We would like to stress the importance of the pressure fluctuations in the growth and self-sustaining of this kind of turbulence. Even though the time average of the related terms vanish from the velocity fluctuation equations, they can be seen as the main engine for isotropizing the turbulence, hence disappearing 
in time average when isotropy is acheived.

Complementary work is needed, in the laboratory or numerically, in order to seek whether the two particular values or the critical Rossby  number matching the experimental results on the Couette-Taylor experiment are genuine to that system or are of more "universal" value.

\begin{acknowledgements}

The author would like to thank Jean-Paul Zahn, Olivier Dauchot,  B\'ereng\`ere Dubrulle, Fran\c{c}ois Daviaud, Sanford Davis, Jeffery Cuzzi, Richard Young, Adriane Steinacker, Franck Hersant and Jean-Marc Hur\'e for stimulating discussions and support. This work was supported by the Programme National de Physique Stellaire (PNPS), the Commissariat \`a l'Energie Atomique  and by a Research Associateship from the National Research Council / National Academy of Sciences.

\end{acknowledgements}

\bibliographystyle{aa}
\bibliography{nonlin}

\appendix

\section{Turbulent fluctuations equations}
\label{ReynoldsStress}

{ The turbulent transport equations are given here in cylindrical coordinates ($r,\phi,z$). Viscous diffusion is omitted}

\bea &\partial_{t}&\overline{{u_{r}^{'}}^{2}}
   +\overline{u_{r}}{\partial_{r} \overline{{u_{r}^{'}}^{2}}} +
   \overline{u_{\phi} \over r} {\partial_{\phi}
   \overline{{u_{r}^{'}}^{2}}} + \overline{u_{z}}{\partial_{z}
   \overline{{u_{r}^{'}}^{2}}} - 2 \overline{u_{\phi} \over r}
   \overline{u_{r}^{'}u_{\phi}^{'}} = \nonumber\\ 
  &-&2 \left( 
   \overline{{u_{r}^{'}}^{2}} {\partial_{r} \overline{u_{r}}} +
   \overline{u_{r}^{'}u_{\phi}^{'}} \left( {\partial_{\phi}
   \overline{u_{r}} \over r} - {\overline{u_{\phi}} \over r} \right) +
   \overline{u_{r}^{'}u_{z}^{'}} {\partial_{z} \overline{u_{r}}} \right) \nonumber\\
   &-& 2 \overline{ u_{r}^{'} \left( {1 \over r}
   \partial_{r}(r {u_{r}^{'}}^{2}) + {1 \over
   r}\partial_{\phi}(u_{r}^{'}u_{\phi}^{'}) -{{u_{\phi}^{'}}^{2}\over r}
   + \partial_{z}(u_{r}^{'}u_{z}^{'}) \right) }\nonumber\\ 
   &+&2 {1\over \rho} \overline{p^{'} \partial_{r}u_{r}^{'}} - 2 {1 \over
   \rho}\overline{ \partial_{r}(p^{'} u_{r}^{'}) }\nonumber \\   
 &-& \nu \Bigg( 2 \overline{\left( \partial_{r}只_{r}匆{'} \right)^{2}}    + \overline{\left( \partial_{r}只_{\phi}匆{'} \right)^{2}} +    \overline{\left( \partial_{r}只_{z}匆{'} \right)^{2}}    \nonumber\\
&& \;\;\;\;\; +\overline{\left( {1 \over r}\partial_{\phi}只_{r}匆{'} - {    u_{\phi}^{'} \over r}\right)^{2}} +\overline{\left(    \partial_{z}只_{r}匆{'} \right)^{2}} \Bigg)
 \label{fullvr}
\eea

\bea
\partial_{t}\overline{{u_{\phi}^{'}}^{2}}
   &+& \overline{u_{r}}{\partial_{r} \overline{{u_{\phi}^{'}}^{2}}}
   + \overline{u_{\phi} \over r} {\partial_{\phi} \overline{{u_{\phi}^{'}}^{2}}}
   + \overline{u_{z}}{\partial_{z} \overline{{u_{\phi}^{'}}^{2}}}
   + 2 \overline{u_{\phi} \over r} \overline{u_{r}^{'}u_{\phi}^{'}}  =\nonumber
\\
    &-& 2 \left( \overline{u_{r}^{'}u_{\phi}^{'}}
   {\partial_{r} \overline{u_{\phi}}} + \overline{{u_{\phi}^{'}}^{2}}
   \left( {\partial_{\phi} \overline{u_{\phi}} \over r} +
   {\overline{u_{r}} \over r} \right) + \overline{u_{\phi}^{'}u_{z}^{'}}
   {\partial_{z} \overline{u_{\phi}}} \right) \nonumber \\
   &-& 2 \overline{ 
   u_{\phi}^{'} \left(
   {1 \over r} \partial_{\phi}({u_{\phi}^{'}}^{2})  +  \partial_{r}(u_{r}^{'}u_{\phi}^{'})
   + 2 {u_{r}^{'}u_{\phi}^{'} \over r} + \partial_{z}u_{\phi}^{'}u_{z}^{'}
   \right)
   }\nonumber\\
   &+&2 {1 \over \rho} \overline{ p^{'} \left( {1 \over r}\partial_{\phi} u_{\phi}^{'} \right)}
   - 2 {1 \over \rho} \overline{ \left( {1 \over r}\partial_{\phi}(p^{'} u_{\phi}^{'}) \right)} \nonumber \\    
&-& \nu \Bigg( \overline{\left( \partial_{r} u^{'}_{\phi}\right)^{2}}    +{1 \over r^{2}}\overline{\left( \partial_{\phi} u^{'}_{z}\right)^{2}}    +\overline{\left( \partial_{z} u^{'}_{\phi}\right)^{2}}   \nonumber\\
 && \;\;+2\overline{\left( {1 \over r} \partial_{\phi} u^{'}_{\phi} + {u^{'}_{r}\over r} \right)^{2}}    +{\overline{\left( {1 \over r} \partial_{\phi} u^{'}_{r} + {u^{'}_{\phi}\over r} \right)^{2}}}    \Bigg) 
\label{fullvphi}
\eea

\bea
  \partial_{t}\overline{{u_{z}^{'}}^{2}}
   &+& \overline{u_{r}}{\partial_{r} \overline{{u_{z}^{'}}^{2}}}
   + \overline{u_{\phi} \over r} {\partial_{\phi} \overline{{u_{z}^{'}}^{2}}}
   + \overline{u_{z}}{\partial_{z} \overline{{u_{z}^{'}}^{2}}}
   =  \nonumber\\
   &-&2
   \left(
   \overline{u_{r}^{'} u_{z}^{'}} {\partial_{r} \overline{u_{z}}} 
   + \overline{u_{z}^{'}u_{\phi}^{'}} {\partial_{\phi} \overline{u_{z}} \over r}
   + \overline{{u_{z}^{'}}^{2}} {\partial_{z} \overline{u_{z}}}
   \right) \nonumber \\
   &-& 2 \overline{
   u_{z}^{'} \left(
   {1 \over r} \partial_{r}(r u_{r}^{'} u_{z}^{'})  +  {1 \over r}\partial_{\phi}(u_{\phi}^{'}u_{z}^{'})
   + \partial_{z}{u_{z}^{'}}^{2}
   \right)
   }\nonumber\\
   &+&2 {1 \over \rho} \overline{p^{'} \partial_{z}u_{z}^{'}}  
   - 2 {1 \over \rho}\overline{ \partial_{z}(p^{'} u_{z}^{'}) } \nonumber\\   
& -& \nu \Bigg( 2 \overline{\left( \partial_{z} u^{'}_{z}\right)^{2}}    +{1 \over r^{2}}\overline{\left( \partial_{\phi} u^{'}_{z}\right)^{2}}    +\overline{\left( \partial_{r} u^{'}_{z}\right)^{2}}   \nonumber\\ 
&& \;\;\;\;\; \;\;\;\;\;+ \overline{\left( \partial_{z} u^{'}_{r} \right)^{2}}    +{\overline{\left( \partial_{z} u^{'}_{\phi} \right)^{2}}}    \Bigg)
\label{fullvz}
\eea

\end{document}